# Suppression of Cross-Field Transport of a Passive Scalar in Two-Dimensional Magnetohydrodynamic Turbulence

P.H. Diamond and A.V. Gruzinov
*Department of Physics 0319*
*University of California, San Diego*
*La Jolla, CA 92093-0319, USA*

## Abstract

The theory of passive scalar transport in two dimensional turbulent fluids is generalized to the case of 2D MHD. Invariance of the cross correlation of scalar concentration and magnetic potential produces a novel contribution to the concentration flux. This pinch effect is proportional to the mean potential gradient, and is shown to drastically reduce transport of the passive scalar across the mean magnetic field when $R_m B_0^2 / 4\pi\rho \langle \tilde{v}^2 \rangle > 1$. Transport parallel to the mean magnetic field is unchanged. Implications for models of transport in turbulent magnetofluids are discussed.

PAC NOS.  47.25.Jn, 47.65.+a

Recent work on the theory of magnetic potential (flux) diffusion in two dimensions[1] and the $\alpha$-effect driven dynamo in three dimensions[2] has focused on the effects of small scale magnetic fluctuations. The thrust of the research pursued by several groups is that Lorentz forces produced by rapidly amplified small scale magnetic fields strongly "back-react" on the fluid dynamics, thus inducing a marked departure of mean field evolution from kinematic predictions. Thus, in two dimensions the magnetic potential diffusivity is determined by the competition between small scale hydrodynamic and magnetic energies,[3] i.e. $\eta_{eff} \sim \langle \tilde{v}^2 \rangle - \langle \tilde{B}^2 \rangle$. Similarly, the $\alpha$-effect is set by the imbalance of fluid helicity and magnetic hyper-helicity,[4] i.e. $\alpha \sim \langle \underline{\tilde{v}} \cdot \underline{\tilde{\omega}} \rangle - \langle \underline{\tilde{B}} \cdot \underline{\tilde{J}} \rangle$. The magnitudes of small scale magnetic energy and hyper-helicity are constrained by topological conservation laws which govern magnetic field evolution. Specifically, in two dimensions, conservation of mean square magnetic potential directly ties the magnetic fluctuation energy to the net spatial transport of mean magnetic flux. Similarly, in three dimensions, conservation of magnetic helicity directly relates the magnetic fluctuation hyper-helicity $\langle \underline{\tilde{B}} \cdot \underline{\tilde{J}} \rangle$ to the net mean-field $\alpha$-effect. In both cases, the mean-field transport coefficient in question (i.e. the flux diffusivity in 2D, $\alpha$-coefficient in 3D) is reduced by a factor $\left(1 + R_m B_0^2 / 4\pi\rho \langle \tilde{v}^2 \rangle\right)^{-1}$ relative to the analogous kinematic prediction. Hereafter, the mean magnetic field-induced suppression factor $\left(1 + R_m B_0^2 / 4\pi\rho \langle \tilde{v}^2 \rangle\right)^{-1}$ is written as $(1+R)^{-1}$, where $R = R_m B_0^2 / 4\pi\rho \langle \tilde{v}^2 \rangle$. The notation is standard, with $R_m$ the magnetic Reynolds number. It follows that inhibition of magnetic flux diffusion and saturation of the dynamo process occur for mean field magnetic energies drastically <u>below</u> the equipartition level (i.e., for $B_0^2 / 4\pi \gtrsim \rho \langle \tilde{v}^2 \rangle / R_m$). Note that for $R > 1$, $D_{eff} = D_K / (1+R)$ scales proportional to the collisional resistivity, *even if the turbulence correlation time is short relative to resistive diffusion times,* i.e. $\tau_c \eta k^2 \ll 1$. This reflects the "freezing in" property of magnetic potential dynamics. Such results call for a re-evaluation of the conventional wisdom developed using kinematic mean-field electrodynamics.[5] In particular, the impact of the novel mechanism for saturation of the dynamo at modest field levels upon predictions derived from conventional dynamo theory should be assessed.

From the discussion above, *one is naturally lead to wonder about the effects of small scale magnetic fields on turbulent transport and mixing in MHD fluids, in general.* This question is obviously relevant to understanding the origin and profile of the solar differential rotation and the structure of the solar convection zone as well as to other problems in astrophysical fluid dynamics. Note that recent analysis of helioseismological data suggests that the solar convection zone is enmeshed by a network of small scale magnetic flux tubes with significant volumetric packing fraction,[6] thus suggesting that the convection zone should be viewed as an MHD fluid. The dynamics of each are governed by turbulent transport of momentum and heat, respectively. As small scale magnetic fluctuations are known to modify the effective (eddy) viscosity coefficients, correlation times and spectra of MHD turbulence, it is indeed reasonable to explore the effect of magnetic turbulence on turbulent transport in magnetofluids. Also, understanding the dynamo in a Reversed Field Pinch plasma requires a theory of the turbulent transport of magnetic helicity.[7] As turbulent momentum and heat transport in MHD fluids are exceedingly complex phenomena, we first seek to identify and understand the dynamics of a simple prototypical model. The most obvious such "hydrogen atom problem" is that of transport of a passive scalar in 2D magnetohydrodynamics (MHD), which is a straightforward generalization of the classic paradigm of passive scalar convection by a turbulent fluid.[8] This problem is contained within the simple, familiar model consisting of the magnetic potential, vorticity and passive scalar concentration advection equations in two dimensions, i.e.

$$\frac{\partial \psi}{\partial t} + \underline{\nabla}\phi \times \hat{z} \cdot \underline{\nabla}\psi = \eta \nabla^2 \psi, \tag{1a}$$

$$\frac{\partial}{\partial t}\nabla^2\phi + \underline{\nabla}\phi \times \hat{z} \cdot \underline{\nabla}\nabla^2\phi - \nu\nabla^2\nabla^2\phi = \underline{\nabla}\psi \times \hat{z} \cdot \underline{\nabla}\nabla^2\psi, \tag{1b}$$

$$\frac{\partial}{\partial t}c + \underline{\nabla}\phi \times \hat{z} \cdot \underline{\nabla}c - D_0 \nabla^2 c = 0. \tag{1c}$$

$\phi$ and $\psi$ are the velocity stream function and magnetic potential, respectively. Thus fluid velocity $\underline{v} = \underline{\nabla}\phi \times \hat{z}$ and magnetic field $\underline{B} = \underline{\nabla}\psi \times \hat{z}$. Here the collisional resistivity is $\eta$, the kinematic viscosity is $\nu$ and the collisional diffusivity of the scalar concentration is $D_0$. Hereafter, we take $\eta = \nu$ for simplicity, but allow $D_0 \neq \eta$. This system has the familiar quadratic conserved (up to diffusive dissipation) quantities, which are total energy $\int d^2x \left((\underline{\nabla}\phi)^2 + (\underline{\nabla}\psi)^2\right)/2$, mean square magnetic helicity $\int d^2x (\psi^2)$, and cross-helicity $\int d^2x (\underline{\nabla}\phi \cdot \underline{\nabla}\psi)$. It also has two other quadratic inviscid invariants, namely $\int d^2x (c^2/2)$ and the (important) concentration-magnetic potential (C-MP) correlation $\int d^2x\, c\psi$. Note that the conservation of C-MP correlation is a consequence of the fact that both $c$ and $\psi$ are conserved along fluid element trajectories. Thus $c = c(\phi)$, $\psi = \psi(\phi)$ so $c = c(\psi)$.

For a given turbulent velocity field, a standard approach to the determination of the passive scalar flux $\underline{\Gamma}$ is to proceed via a quasi-linear closure. Specifically, we calculate the mean field flux of scalar concentration as

$$\underline{\Gamma} = \left\langle (\underline{\nabla}\phi \times \hat{z})c \right\rangle \tag{2a}$$

and approximate $\underline{\Gamma}$ as

$$\underline{\Gamma} = \left\{ \left\langle \underline{\nabla}\phi\, c^{(1)} \right\rangle + \left\langle \underline{\nabla}\phi^{(1)} c \right\rangle \right\} \times \hat{z} \tag{2b}$$

where $c^{(1)}$ and $\phi^{(1)}$ are determined from Eqns. (1c) and (1b) respectively, assuming finite correlation time for the MHD turbulence. Thus, taking $\langle c \rangle$ and $\langle \psi \rangle$ to refer to the mean concentration profile and magnetic flux (where $\langle \underline{B} \rangle = \underline{\nabla}\langle \psi \rangle \times \hat{z}$), it follows that the responses $c^{(1)}_{\underline{k}}$ and $\phi^{(1)}_{\underline{k}}$ are:

$$c_{\underline{k}}^{(1)} = -\tau_{c\underline{k}}(\underline{\nabla}\phi)_{\underline{k}} \times \hat{z} \cdot \underline{\nabla}\langle c \rangle, \tag{3a}$$

$$\nabla^2 \phi_{\underline{k}}^{(1)} = \tau_{c\underline{k}}\left[\underline{\nabla}\langle\psi\rangle \times \hat{z} \cdot \underline{\nabla}\left(\nabla^2\psi^{(1)}\right)_{\underline{k}} + (\underline{\nabla}\psi)_{\underline{k}}^{(1)} \times \hat{z} \cdot \underline{\nabla}\nabla^2\langle\psi\rangle\right]. \tag{3b}$$

Contributions from the vorticity advection nonlinearity vanish, as $\langle \phi c \rangle_{\underline{k}} = 0$. Neglecting terms of order $1/k^2 L_\psi^2$, where $L_\psi$ is the characteristic scale length of the mean magnetic potential, one then finds:

$$\phi_{\underline{k}}^{(1)} = \tau_{c\underline{k}}(\underline{\nabla}\psi)_{\underline{k}} \times \hat{z} \cdot \underline{\nabla}\langle\psi\rangle \tag{3c}$$

Taking the $\phi$, $\psi$ and $c$ fields to be isotropic then yields:

$$\underline{\Gamma} = -\sum_{\underline{k}} \tau_{c\underline{k}}\left[\left\langle(\underline{\nabla}\phi)^2\right\rangle_{\underline{k}} \underline{\nabla}\langle c \rangle - \langle\underline{\nabla}c\cdot\underline{\nabla}\psi\rangle_{\underline{k}} \underline{\nabla}\langle\psi\rangle\right] \tag{4a}$$

$$= -D_K \underline{\nabla}\langle c \rangle + V\underline{\nabla}\langle\psi\rangle \tag{4b}$$

Note that the total scalar concentration flux consists of the usual diffusive piece $-D_K \underline{\nabla}\langle c \rangle$, (with the usual kinematic diffusion coefficient $D_K = \sum_{\underline{k}}\langle\underline{\nabla}\phi^2\rangle_{\underline{k}} \tau_{c\underline{k}}$) and a novel magnetic potential gradient driven flux $V\underline{\nabla}\langle\psi\rangle$ (with pinch velocity coefficient $V = \sum_{\underline{k}}\langle\underline{\nabla}c\cdot\underline{\nabla}\psi\rangle_{\underline{k}} \tau_{c\underline{k}}$). Since the $\underline{\nabla}\langle\psi\rangle$ flux vanishes along contours of constant mean magnetic potential, *only the cross-field flux is renormalized by the pinch term*, i.e. $\underline{\Gamma}_\| = \underline{\Gamma}_{\|,K}$, the kinematic result.

The readily evident proportionality of the pinch $V$ to the correlation $\langle\underline{\nabla}c\cdot\underline{\nabla}\psi\rangle$, which is obviously related to the C-MP correlation, gives a clue as to the importance of the (topological) conservation of mean-square magnetic potential for passive scalar transport.

The conservation of C-MP correlation in the presence of mean profiles $\langle c \rangle$ and $\langle \psi \rangle$ implies:

$$\frac{\partial}{\partial t}\langle c\psi \rangle + \langle \underline{v}\psi \rangle \cdot \underline{\nabla}\langle c \rangle + \langle \underline{v}c \rangle \cdot \underline{\nabla}\langle \psi \rangle = -(\eta + D_0)\langle \underline{\nabla} c \cdot \underline{\nabla}\psi \rangle. \tag{5a}$$

Noting that by definition $\langle \underline{v}\psi \rangle = -\eta_{eff}\underline{\nabla}\langle \psi \rangle$, $\langle \underline{v}\tilde{c} \rangle_\perp = -D_{\perp eff}\underline{\nabla}\langle c \rangle$ and requiring stationarity, the following relation between the evolution of C-MP correlation, transport of magnetic potential and scalar concentration, and the collisional dissipation of these quantities follows directly

$$\left(\eta_{eff} + D_{\perp eff}\right)\underline{\nabla}\langle c \rangle \cdot \underline{\nabla}\langle \psi \rangle = (\eta + D_0)\langle \underline{\nabla} c \cdot \underline{\nabla}\psi \rangle. \tag{5b}$$

Here $\eta_{eff} = D_K/[1+R]$, and recall $D_K$ is the kinematic turbulent diffusion coefficient. Thus, the correlation $\langle \underline{\nabla} c \cdot \underline{\nabla}\psi \rangle$ may be obtained from Eqn. (5b) and substituted into Eqn. (4) for $V$, which yields

$$\underline{\Gamma} = -D_K \underline{\nabla}\langle c \rangle + \tau_c \frac{\left(\underline{\nabla}\langle c \rangle \cdot \underline{\nabla}\langle \psi \rangle\right)\left(D_{\perp eff} + \eta_{eff}\right)\underline{\nabla}\langle \psi \rangle}{(\eta + D_0)} \tag{6}$$

Now, we are concerned with cross-field transport only, so we take $\underline{\nabla}\langle c \rangle$ parallel to $\underline{\nabla}\langle \psi \rangle$. Thus

$$\Gamma_\perp = -D_K \nabla_\perp \langle c \rangle + \tau_c \frac{\left(\underline{\nabla}\langle \psi \rangle\right)^2}{(\eta + D_0)}\left(D_{\perp eff} + \eta_{eff}\right)\nabla_\perp \langle c \rangle. \tag{7}$$

Use of the definition $\Gamma_\perp = -D_{\perp eff}\nabla_\perp \langle c \rangle$ and some straightforward manipulation then finally reveal the effective cross field diffusivity as predicted by mean field theory to be

$$D_{\perp eff} = \frac{D_K}{(1+R)} \left[ \frac{1 + \frac{D_0}{\eta+D_0} R}{1 + \frac{\eta}{\eta+D_0} R} \right]. \tag{8}$$

The form of $D_{\perp eff}$ obtained above is non-trivial. For $\eta = D_0$ which corresponds to Prandtl number (as well as magnetic Prandtl number) unity, $D_{\perp eff} = D_K(1+R)^{-1}$. Hence, the cross-field diffusivity is suppressed, in comparison to the kinematic prediction, by the same factor as the magnetic potential diffusivity is reduced. This may be understood by noting that in the limit of $\nu = \eta = D_0$, $c = c(\phi)$ and $\psi = \psi(\phi)$ so $c = c(\psi)$ *on all inertial scales*. As magnetic potential transport is suppressed for $R = B_0^2 R_m / \rho \langle \tilde{v}^2 \rangle > 1$, and the scalar concentration is frozen into the magnetic potential, it is not surprising that the cross-field scalar diffusivity is similarly quenched. This interpretation is supported by the observation that for $D_0 \gg \eta$, $D_{\perp eff} \to D_K$, i.e. the kinematic result is recovered. The reversion to simple kinematics in this limit follows from the breakdown of the relation $c = c(\psi)$ due to the disparity between $D_0$ and $\eta$. Simply put, since the passive scalar concentration $c$ is *not* frozen into the fluid, the fact that magnetic potential *is* frozen into the fluid (thus suppressing magnetic diffusion) is dynamically irrelevant. Curiously, for $D_0 \to 0$, $D_{\perp eff} = D_K/(1+R)^2$. In this limit, cross field passive scalar transport is quenched (for $R > 1$) even more strongly than magnetic potential diffusion is. Care should be taken in drawing conclusions from this result, however, since $D_0 \to 0$ may not be compatible with a fully stationary state. Finally, the absence of $D_\parallel$ renormalization may be understood by realizing that the suppression of magnetic flux transport is due to a kind of "elastic memory" of magnetic field lines in a high magnetic Reynolds number fluid. Thus, a "plucked" field line tends to "snap back" to its original configuration. Hence, the restoring force implied by the elastic memory analogy acts in a direction *perpendicular* to the mean magnetic field. Since $c = c(\psi)$, only the cross field transport is quenched.

The predictions presented above can easily be tested by a simple numerical experiment in which a mean magnetic potential profile $\langle\psi(x)\rangle$ and a circular blob of concentration density $c(\underline{x})$ centered at the origin are initialized in a turbulent 2D magnetofluid. The magnetic potential amplitude should be adjustable, so that cases with $R<1$ and $R>1$ can be compared. For $R<1$, the concentration blob will spread isotropically with $\langle\delta r^2\rangle \sim D_K\tau$. For $R>1$, the blob should develop a pronounced anisotropy with variances along the principal axes given by $\langle\delta y^2\rangle \sim D_K\tau$ and $\langle\delta x^2\rangle \sim D_{\perp eff}\tau \sim D_K\tau/(1+R)$ (for $\eta \sim D_0$). Keep in mind that for $R_m >> 1$ (the case of interest) $\langle B\rangle^2/4\pi << \rho\langle\tilde{v}^2\rangle/2$, so that the mean field can be well below the equipartition level. This is because the dynamically relevant field is the <u>small</u> scale magnetic field, the energy of which is amplified by stretching of $\langle B \rangle$. Here, the parameter $R$ should be thought of as a measure of the strength of small scale magnetic field effects in terms of mean field strength.

The least-defensible approximation made in this paper is that of a single correlation time $\tau_{c\underline{k}}$ for each of the spectra $\langle\phi^2\rangle_{\underline{k}}$, $\langle\psi^2\rangle_{\underline{k}}$ and $\langle c^2\rangle_{\underline{k}}$. In strong MHD turbulence, when inertial range equipartition occurs, it is however reasonable to expect both $\langle\tilde{v}^2\rangle_{\underline{k}}$ and $\langle\tilde{B}^2\rangle_{\underline{k}}$ to have the same correlation time, which will, however, depend on $\underline{B}_{rms}$ and $\underline{B}_0$ (i.e. via the Alfven effect).[9] Since $c$ is a passive scalar, with dynamics determined by $\langle\tilde{v}^2\rangle_{\underline{k}}$ and (indirectly) by $\langle\tilde{B}^2\rangle_{\underline{k}}$, it does indeed appear reasonable to hypothesize a single correlation time for this system. Another assumption made ab-initio here is that the microscopic collision frequency $\nu_c$ of the gas exceeds the (electron) cyclotron frequency $(\nu_c > \Omega_e)$, so that the passive scalar diffusion tensor is invariant with respect to the direction of $\underline{B}_0$. If $\Omega_e > \nu_c$, the collisional flux of certain scalar quantities becomes anisotropic. Specifically, electron heat conduction along the field has diffusivity $\chi_{\|} = V_T^2/\nu_{e,e}$ but $\chi_\perp = \nu_{ee}\rho_e^2$. Moreover, $D_0\nabla^2 T \to \nabla_{\|}\chi_{\|}\nabla_{\|}T + \chi_\perp \nabla_\perp^2 T$, where $\nabla_{\|} = (\underline{B}_0 + \underline{\tilde{B}}\cdot\nabla)/|B|$. Hence, anisotropic transport introduces new non-linearities and a different transport mechanism mediated by magnetic fluctuations, only.[10]

The implications of these results must be discussed keeping in mind the obvious question of the relevance of a 2D MHD model (where $\langle \psi^2 \rangle$ is conserved) to 3D dynamics where magnetic helicity, not mean square potential, is the relevant topological invariant. First, it should be noted that magnetized incompressible MHD fluids are well described the reduced MHD model,[11] where $\langle \psi^2 \rangle$ invariance is broken only by the bending of $B_z$ (i.e. $\partial \langle \psi^2 \rangle / \partial t \sim B_z \langle \tilde{\psi} \partial \hat{\phi} / \partial z \rangle$). Thus, situations of *approximate* "mean" square potential conservation can arise particularly when $\tau_c k_z V_A < 1$ (i.e. turbulence correlation time is shorter than the parallel Alfven transit time). The transport suppression predictions given here are directly relevant to such situations. Moreover, the theory discussed here could be extended to incorporate the effects of Alfvenic radiation in the mean square magnetic potential budget. Second, many aspects of the phenomena discussed here are generic to MHD turbulence. For example, mean flow evolution by turbulent transport is governed by the competition between fluid and magnetic stresses i.e.

$$\frac{\partial}{\partial t} \langle V_i \rangle = -\frac{\partial}{\partial x_j} \left[ \langle \tilde{V}_j \tilde{V}_i \rangle - \frac{\langle \tilde{B}_j \tilde{B}_i \rangle}{4\pi \rho_0} \right] \quad (9)$$

Hence, amplification of $\langle \tilde{B}^2 \rangle$ to equipartition levels will certainly affect (and likely reduce) turbulent momentum transport. This scenario, which closely resembles that of the prototypical model discussed above, is not encompassed by existing theories of solar differential rotation. Indeed, the latter treat only the effects of mean magnetic fields.[12] Finally, the results suggest a re-evaluation of the lore concerning turbulent transport in magnetofluids. Specifically, the effects of small scale magnetic fields on eddy conductivity and mixing length theory should be investigated. Also, greater attention should be given to mechanisms for the spontaneous amplification of magnetic fluctuations, such as the magnetic shearing instability.[13] Here, quasilinear predictions of saturation levels with $\langle \tilde{B}^2 \rangle / 8\pi > \rho \langle \tilde{v}^2 \rangle / 2$ are *intrinsic* to the instability dynamics, thus avoiding the cancellations induced by growth to equipartition.

We thank F. Cattaneo, M.B. Isichenko, S.I. Vainshtein, and E.T. Vishniac for stimulating discussions. This research was supported by the Department of Energy under Grant No. DE-FG03-88ER53275 and NASA under Grant No. NAGW-2418.